\documentclass[preprint,a4paper,12pt]{elsarticle}

\usepackage{amsmath}
\usepackage{amsfonts}
\usepackage{amssymb}
\usepackage{amsthm}

\usepackage{algorithm}
\usepackage[noend]{algpseudocode}
\usepackage{graphicx}
\usepackage{subcaption}
\usepackage{multirow} 
\usepackage{float} 

\theoremstyle{definition}
\newtheorem{theorem}{Theorem}[section]
\newtheorem{lemma}[theorem]{Lemma}

\usepackage{xcolor}
\usepackage[colorlinks=true, allcolors=blue]{hyperref}
\def\RR{\mathbb R}

\def\ZZ{\mathbb Z}

\def\bh{\mathbf h}
\def\bp{\mathbf p}
\def\br{\mathbf r}
\def\bx{\mathbf x}
\def\bu{\mathbf u}
\def\bz{\mathbf z}
\def\bw{\mathbf w}
\def\bT{\mathbf T}
\def\bzero{\mathbf 0}
\newcommand{\hide}[1]{}

\newcommand{\sproofof}[1]{\noindent{\textit{Proof of~{#1}.}}}
\newcommand{\raf}[1]{(\ref{#1})}
\newcommand{\ORAD}{\OMKD}
\newcommand{\lbORAD}{\ensuremath{\mathsf{lb\OMKD}}}

\newcommand{\OKP}{\ensuremath{\mathsf{OKP}}}

\newcommand{\OMKD}{\ensuremath{\mathsf{OMKD}}}

\newcommand{\OMK}{\ensuremath{\mathsf{OMK}}}
\newcommand{\OMdK}{\ensuremath{\mathsf{OMdK}}}

\newcommand{\MdOMKD}{\ensuremath{\mathsf{MdOMKD}}}
\newcommand{\CR}{\ensuremath{\mathsf{CR}}}
\newcommand{\obj}{\ensuremath{\mathsf{obj}}}

\begin{document}
	
\begin{frontmatter}

\title{Online Multiple Resource Allocation Problems with Departures via the Primal-Dual Approach}

\author[inst1]{Yusuf Amidu}

\author[inst2]{Khaled Elbassioni}

\author[inst1]{Adriana F. Gabor\corref{cor1}}
\cortext[cor1]{Corresponding author.}
\ead{adriana.gabor@ku.ac.ae}

\affiliation[inst1]{
	organization={Department of Mathematics, Khalifa University of Science \& Technology},
	city={Abu Dhabi},
	postcode={PO Box 127788}, 
	country={United Arab Emirates}
}

\affiliation[inst2]{
	organization={Department of Computer Science, Khalifa University of Science \& Technology},
	city={Abu Dhabi},
	postcode={PO Box 127788}, 
	country={United Arab Emirates}
}


\begin{abstract}
	In this paper we propose primal-dual algorithms for different variants of the online resource allocation problem with departures. In the basic variant, requests (items) arrive over time to a set of resources (knapsacks) and upon arrival, the duration of time a request may occupy a resource, the demand and reward if the request can be granted, become known. 
The goal of the algorithm is to decide whether to accept/reject a request upon arrival and to which resource to allocate it such that the reward obtained over time is maximized. Under some mild assumptions, we show that the proposed primal-dual algorithm achieves a competitive ratio of $O\big(\log(\bar\theta^{\max}\cdot\bar d^{\max})\big)$, where $\bar \theta^{\max}$ is the maximum value density fluctuation ratio and $\bar d^{\max}$ is the maximum duration fluctuation ratio. We prove similar results for two other variants, namely, one with an additional load balancing constraint, and the multi-dimensional variant where an admitted request consumes capacity on multiple resources. Our results show that the primal-dual approach offers a simple, unified framework for obtaining competitive ratios comparable to those previously obtained via threshold policies known for these problems. Additionally, we show that this framework allows us to incorporate additional constraints, such as load-balancing constraints, without sacrificing the competitive ratio. 
\end{abstract}

\begin{keyword}
	  Online Knapsack Problems \sep Primal-Dual Algorithms \sep Competitive Ratio
\end{keyword}
	
\end{frontmatter}

\section{Introduction}
Resource allocation is traditionally viewed as an optimization problem, where a set of requests must be distributed across a set of resources with limited capacity to achieve the most efficient use of these resources. The typical goals are to minimize total costs, processing time, or conflicts, or to maximize overall profit, efficiency, or compatibility \cite{datta2007multi}. In the online version, requests arrive over time and a decision of acceptance/rejection has to be made for each arriving request, such that the capacity of resources is not exceed while some objective is minimized/maximized.
The online resource allocation problem has a wide range of  applications across several fields, including computer resource allocation \cite{kurose1989microeconomic}, cloud and edge computing \cite{lucier2013efficient, zheng2016online}, online admission control \cite{buchbinder2009design}, virtual switch routing \cite{paris2016online}, and management of distributed energy resources in smart grids \cite{alinia2019online, alinia2020online, alinia2018competitive, sun2020orc}. 

The simplest form of the online resource allocation problem is the online knapsack problem (\OKP) where one needs to decide whether to accept/reject requests (items) arriving over time to a single resource (knapsack). Once a decision has been taken, it cannot be revoked. Each item is characterized by a reward and a size (weight). The objective is to  maximize the total value of accepted items without exceeding the knapsack’s capacity. Several extensions have been explored over the years, such as:
  (i) the Online Multiple Knapsack (\OMK) problem, which involves multiple knapsacks and requires determining both whether to accept an item and which knapsack to assign it to; (ii) the Online Multi-dimensional Knapsack (\OMdK) problem, where each item has a vector of resource requirements, and decisions must ensure multi-dimensional capacity constraints are met; and (iii) the Online Multiple Knapsack with Departures (\OMKD) problem, where items that arrive also leave the system after a limited duration of stay.

 The focus of this paper is on developing simple primal-dual algorithms for Online Multiple Knapsack with Departures (\OMKD) and its multidimensional variant, with competitive ratios comparable to the ones obtained via threshold algorithms in \cite{yang2021competitive,sun2022online}. This yields a simple and unified framework to reconstruct the results in these two papers which were obtained using more complicated arguments, and answers an open question raised in~\cite{sun2022online} about the possibility of using the primal-dual approach for this purpose. We also show that the algorithm and its analysis can be easily extended to address additional constraints for which threshold algorithms do not seem straightforward to extend, such as \OMKD\ with additional load balancing constraints.  

The remainder of this paper is structured as follows. In section \ref{section: related work} we review some related work. Section \ref{OMRAD} contains the description of the \OMKD\ and the variants of interest, as well as a comparison of our results with the ones in the literature. Section \ref{alg_result} presents the primal-dual algorithm for \OMKD\ and its analysis. Sections \ref{sec:load} and \ref{sec:multi} present the results for the load balancing and multi-dimensional versions, respectively. Proofs omitted due to tack of space can be found in the appendix.

\section{Related work} \label{section: related work}
The Online Knapsack Problem (\OKP) has been widely examined within the framework of competitive analysis  \cite{marchetti1995, chakrabarty2008online, zhang2017optimal, yang2021competitive}. In this setting, it is desired to to design online algorithms which achieve a small competitive ratio, defined as the worst-case ratio of objective values obtained by the offline algorithm in hindsight and the online algorithm.
\cite{marchetti1995} shows that without additional assumptions on the setup information, no competitive online algorithms can be designed. For the case of infinitesimal item sizes and fluctuation (ratio between maximum and minimum) of the item value density bounded by $\bar\theta$, \cite{chakrabarty2008online} proposes a threshold-based algorithm with a competitive ratio $\ln(\bar\theta)+1$ for the \OKP\ and \OMK\ and shows that no online algorithm can achieve a lower competitive ratio. \cite{zhang2017optimal} extends the algorithm to Multi-dimensional Knapsack problem (\OMdK), obtaining a competitive ratio of $O(m\log(\bar\theta))$, where $m$ is the number of dimensions of the knapsack. Based on a novel threshold function, \cite{yang2021competitive} improves the competitive ratio to $O(\log(\bar\theta\cdot\frac{\sum_kC_k}{\min_k C_k}))$.

To the best of our knowledge, \cite{zhang2017optimal} is the first paper to study the Online Multiple Knapsack problem with departures (\OMKD). By treating each time-slot in a time horizon $T$ as a different dimension of a knapsack, the algorithm in~\cite{zhang2017optimal} obtains a competitive ratio of $O(|T|\log(\bar\theta))$, which can further be improved to $O(\log(|T|\bar\theta\cdot\frac{\sum_kC_k}{\min_k C_k}))$ based on \cite{yang2021competitive}. By exploiting the fact that departures of requests lead to a local dependency between requests in different time slots, \cite{sun2022online} reduces the competitive analysis over the horizon $T$ to the analysis of smaller instances, leading to a threshold algorithm with a competitive ratio $O(\log (\bar d \cdot\bar\theta))$ where  $\bar d$ is the maximum duration fluctuation (the ratio between the maximum and minimum duration of a request). 
For the Multi-dimensional Online Multiple Knapsack with Departures (\MdOMKD), \cite{sun2022online} proposes an algorithm that achieves a competitive ratio of $O(\log(\bar d\cdot\bar\rho\cdot\frac{\sum_kC_k}{\min_k C_k}))$, where $\bar d$ and $\bar\rho$ are the maximum fluctuations in durations and value densities, respectively, and shows that no online algorithm can achieve a smaller competitive ratio.  

The potential of the primal dual framework in designing online algorithms, especially for packing and covering problems, is widely recognized \cite{buchbinder2009design}. While it is known that the the framework can be adapted to design an online algorithm for the \OKP\ with $O(\log(\bar\theta))$ competitive ratio, under the assumptions in \cite{chakrabarty2008online}, the possibility of extending it to obtain integer solutions for \OMKD\ with an order-optimal a competitive ratio was posed as an open question in \cite{sun2022online}. 

In this paper, we close this gap by presenting a primal-dual algorithm based on an exponential updating scheme for the dual that achieves a competitive ratio comparable to the one obtained by threshold algorithms. The algorithm can be easily adapted to address the multi-dimensional variant and a novel load balancing variant. 

\section{The Online Multiple Resource Allocation Problem with Departures 
}\label{OMRAD}

\subsection{The Basic Model}\label{online}

We borrow much of our notation from~\cite{sun2022online}. In the basic  variant of the problem, denoted by \OMKD, we are given a set $K$ of $m$ resources and a set \(N\) of requests. Each resource $k\in K$  has a capacity $C_k\ge 0$, and each request $n\in N$ has an associated reward $v_{nk}\ge0$ that would be realized if a demand (or weight) of $w_{nk}\ge0$ is  allocated for a duration of $d_{nk}$ time slots on resource $k$, starting with time $s_{nk}$. Let $T$ be the time horizon and denote by $a_n\in T$ the arrival time for request $n$ and by $T_{nk} = \{s_{nk}, s_{nk}+1, ..., s_{nk}+d_{nk}-1 \}\subseteq T$ the time interval request $n$ demands to occupy resource $k$. We assume the standard online model where the requests arrive over time; upon the arrival of  request $n$ (at time $a_n$), its corresponding information is revealed, namely, the demand $w_{nk}$, the duration $T_{nk}$, and the reward $v_{nk}$ on each resource $k$. Decision to accept/reject a request is made immediately on arrival of each request without any prior information of the  requests arriving in the future. The aim is to design an algorithm that generates a feasible allocation of requests to resources  such that the total rewards of allocated requests is maximized without violating the capacity constraints. 

\subsection{Variants}
We consider two more variants of the online problem. In the first one, which we call the {\it load-balancing} variant and denote by \lbORAD, requests arrive in batches and we are given an upper bound   $q_k\in  \mathbb{Z}^+$, for each resource $k\in K$. As a form of balancing the load among resources, we consider the additional restriction that, when a batch of requests arrives, no more than $q_k$ requests can be allocated to resource $k$.

In the {\it multi-dimensional multiple knapsack} variant, denoted by \MdOMKD, the resource consumption of a request $n$ in resource (knapsack) $k$ is modeled as a non-negative vector $\bw_{nk}=:(w_{nkm})_{m\in M_{k}}$ over a set of dimensions $M_{k}$. The capacity of $m$th dimension of knapsack $k$ is denoted by $C_{km}$. Once a request $n\in N$ is accepted (or admitted), it consumes an amount of $w_{nkm}$ on {\it each} dimension $m\in M_{k}$ of resource $k$ over the period $T_{nkm}$ of length $d_{nkm}$, and a reward of $v_{nk}$ is received. 

\subsection{Supplementary Notations and Assumptions}\label{assump}

Following~\cite{sun2022online}, we denote by $\theta_{nk}=\frac{v_{nk}}{w_{nk}d_{nk}}$ the {\it value density} of request $n\in N$ in resource $k\in K$ in a given \OMKD\ or \lbORAD\ instance. For \MdOMKD, we define the value density of request $n$ in resource $k$ as $\rho_{nk}=\frac{v_{nk}}{\sum_{m\in M_k}w_{nkm}d_{nkm}}$, and the demand {\it total} fluctuation as $\xi_{nk}=\frac{\sum_{m\in M_k}w_{nkm}}{\min_{m\in M_k:w_{nkm}\ne0}w_{nkm}}$.

For a vector $\br=(r_1,\ldots,r_\ell)\in\RR_+^{\ell}$, we write $r^{\max}=\max_{i}r_i$, $r^{\min}=\min_{i:r_i>0}r_i$, and denote by $\bar r=\frac{r^{\max}}{r^{\min}}$ the fluctuation in $\br$. For instance, $\theta_k^{\max}=\max_{n\in N}\theta_{nk}$ is the maximum value density of a request on resource $k$, 
$\bar d_k=\frac{d_k^{\max}}{d_k^{\min}}$, etc. We also let $\bar d^{max}=\max_{k\in K} \bar d_k$, 
$\theta^{max}=\max_{k\in K} \theta_k$, etc.

 \medskip
 
 The following assumptions are made regarding the input of the problem~\cite{YZHST21,sun2022online}.

\textbf{(A1)} The value density of each request $n$ in resource $k$ is bounded within an interval known to the online algorithm, that is, $\theta_{nk}\in[\theta^{\min}_k,\theta^{\max}_k]$ ($\rho_{nk}\in[\rho_k^{\min},\rho_k^{\max}]$ for \MdOMKD), and the fluctuations $\bar\theta_k$ and $\bar\rho_k$ are finite and known to the algorithm. For \MdOMKD, we also assume that the algorithm knows $\xi_k^{\max}$.

\textbf{(A2)} The duration $d_{nk}$ of each request $n$ in resource $k$ is bounded, that is,  $d_{nk} \in [d^{min}_k, d^{max}_k]$ where the fluctuation $\bar d_k$ is finite and known to the online algorithm. 

\textbf{(A3)} The weight  $w_{nk}$ of each request $n$ in resource $k$ is bounded sufficiently below $C_k$, that is, $w_{nk} \leq \epsilon_k \leq C_k,~\forall~ n\in N,~\forall k\in K$, where $\epsilon_k=o(C_k)$. For \MdOMKD, we assume that the weight of each request $n$ in each dimension $m$ of a resource $k$ is bounded sufficiently below $C_{km}$, that is, $w_{nkm}\leq \epsilon_{km} \leq C_{km}~\forall~ n\in N$, where $\epsilon_{km}=o(C_{km})$.


\hide{
The sets and  parameters used for the resource allocation model are described in Table \ref{table_1}.

\begin{table}[H]
\begin{center}
\caption{Sets and parameters used for the online resource allocation model}\label{table_1}
\begin{tabular}{l l l l l l }
\hline \hline
\textbf{Sets}   \\  
\hline 
$N$ & sets of requests where each request $n\in N$\\
$K$  & sets of resources where each resource $k\in K$  \\
\hline 
\textbf{Parameters}   \\  
\hline
$C_k$ & capacity of resource $k$\\
$s_{nk}$ & the start time of request $n$ in resource $k$\\
$d_{nk}$ & the duration of stay of  request $n$ in resource $k$\\
$T_{nk}$ & the available slot of the duration of stay of request $n$ in resource $k$\\
$v_{nk}$ & the reward obtained by assigning request  $n$ to resource $k$\\
$w_{nk}$ & the weight/size of request $n$ in resource $k$  & \\
\hline
\hline \hline
\end{tabular}
\end{center}   
\end{table}
}
\subsection{Our Results and Comparison with Existing Work}\label{main_result}

We show that the primal-dual framework can be used to achieve logarithmic competitive ratios (CR) for the three variants of \ORAD:
\begin{itemize}
    \item For the basic variant ORAD, where $\epsilon_k=O(\frac1{\gamma}) C_k$ and $\gamma=\log(\bar\theta^{\max}\cdot\bar d^{\max})$, we obtain a CR\ of $O(\gamma)$. This matches exactly the bound obtained in \cite{sun2022online} via threshold policies under the same assumptions (where it was also assumed that $\theta^{\min}_k=1$ for all $k\in K$). 
    \item For the load-balancing variant \lbORAD, where $\epsilon_k=O(\frac1{q_k\gamma}) C_k$ and $\gamma=\log(\bar\theta^{\max}\cdot\bar d^{\max})$, we obtain a CR\ of $O(\gamma)$.
    \item For the multi-dimensional variant \MdOMKD, where $\epsilon_{km}=O(\frac1{\gamma_{k}}) C_{km}$ and $\gamma_k=\log(\bar \rho^{\max}\cdot\xi^{\max}\cdot\bar d^{\max})$, we obtain a \CR\ of $O(\gamma)$. Although this seems incomparable to the bound of $O\big(\log(\bar\rho^{\max}\cdot\frac{\sum_kC_k}{C^{\min}}\cdot\bar d^{\max})\big)$ obtained under the same assumptions in~\cite{sun2022online,yang2021competitive}, our bound becomes stronger when the maximum total fluctuation in the demands of any request is not large compared to the total fluctuation in the capacities (that is, when $\max_{n,k}\frac{\sum_{m\in M_k}w_{nkm}}{ w_{nk}^{\min}}$ is much smaller than $\frac{\sum_kC_k}{C^{\min}}$).
\end{itemize}

We note that the aforementioned results in \cite{sun2022online,yang2021competitive} were obtained using threshold-based algorithms whose competitive ratio analysis involved a number of cases and used substantially different arguments for the two variants  ORAD\ and mdORAD. In that sense, it seems interesting that the primal-dual approach offers a simple, unified framework for obtaining competitive ratios comparable to those previously obtained via threshold policies. Additionally, this framework allows one to incorporate additional constraints, such as load-balancing constraints, without sacrificing the competitive ratio. 

We refer the reader to~\cite{buchbinder2009online} for an extensive description of the primal-dual framework for designing competitive online algorithms. In particular, we use ideas from Chapter 14 (Fractional Packing/Covering) in~\cite{buchbinder2009online} for analyzing our algorithms for the three variants, and combine those ideas with the use of {\it strong LP duality} for \lbORAD\ in the spirit of Chapter 10 (Maximizing Ad-Auctions Revenue) in~\cite{buchbinder2009online}.  

\section{The Basic Variant}\label{alg_result} 
\subsection{The Offline Problem and High-level Idea of the Primal-dual Framework}
In the offline version of the problem, one assumes that the information of all requests is revealed upfront. Let $x_{nk}$ be a binary variable indicating whether request $n$ is assigned to resource $k$.
The offline problem can be modeled as the following integer program:  
\begin{align}
(\textit{IP}) \quad &\max_{\mathbf{x}} \quad \sum_{n \in N,k \in K} v_{nk} x_{nk} \nonumber \\
\text{s.t.} \quad & \sum_{k \in K} x_{nk} \leq 1 \quad \forall~ n \in N,\label{eq:assignment_rel_1} \\
& \sum_{n \in N: t \in T_{nk}}w_{nk} x_{nk} \leq C_k \quad \forall~ k \in K,~ \forall~ t \in T, \label{eq:capacity_rel_1} \\
& x_{nk} \in \{0,1\} \quad \forall~ n \in N,~ k \in K. \nonumber
\end{align}
The objective of the above ({\it IP}) is to maximize the total reward that can be obtained from the allocation of requests to resources. Constraints (\ref{eq:assignment_rel_1})  guarantee that each request is assigned to at most one resource. Constraints (\ref{eq:capacity_rel_1}) require that for each resource at any given time, the total demand of requests that are assigned to a resource does not exceed the capacity of the resource. 

Consider the linear programming (LP) relaxation ($P$) obtained by relaxing  $x_{nk}\in [0,1]$ and its dual program ($D$), given by

\vspace{-.1in}
{\centering \hspace*{-10pt}
\scriptsize
\begin{minipage}[t]{.55\textwidth}
  \begin{alignat}{2}
    (P) \quad & \max_{\mathbf{x}} \quad \sum_{n \in N,k \in K} v_{nk} x_{nk} \nonumber \\
 \text{s.t.} \quad & \sum_{k \in K} x_{nk} \leq 1 \quad \forall~ n \in N, \label{eq:assignment_rel11} \\
 & \sum_{n \in N: t \in T_{nk}} w_{nk} x_{nk} \leq C_k \quad \forall~ k \in K,~ \forall~ t \in T, \label{eq:capacity_rel11} \\
 & x_{nk} \geq 0 \quad \forall~ n \in N,~ k \in K. \nonumber
  \end{alignat}
\end{minipage}
\hspace{0.45em}
\vrule width 0.7pt
\hspace*{-10pt}
\hspace{0.45em}
\begin{minipage}[t]{.55\textwidth}
  \begin{alignat}{2}
    (D) \quad \min_{\bu,\bp}&~~ \sum_{n\in N}u_{n}+\sum_{k\in K,t\in T_{nk}}C_kp_{kt}  \nonumber\\ 
\text{s.t}\quad 
 &u_n + w_{nk}\sum_{t\in T_{nk}}p_{kt}\geq v_{nk} \quad \forall~ n\in N,~k\in K,\label{constr_dual: u_n}\\
 & u_{n} \geq 0 \quad \forall~ n\in N,\nonumber\\
 &p_{kt} \geq 0 \quad \forall~k\in K\nonumber. 
  \end{alignat}
\end{minipage}}
\vspace{0.1in}

\noindent where $u_n$ and $p_{kt}$ are the dual variables corresponding to constraints (\ref{eq:assignment_rel11}) and 
(\ref{eq:capacity_rel11}) in the LP relaxation, respectively. We can interpret $p_{kt}$ as the price for using time slot $t$ of resource $k$ and $u_n$ as the utility of request $n$ (i.e., the reward minus the total price paid for the resource used by $n$).

Let $\bx^*$ and $(\bu^*, \bp^{*})$ represent some optimal solutions to ($P$) and ($D$), respectively. From the complementary slackness conditions, we obtain \\

 (C1) If $x_{nk}^*>0$ then  $u_n^*= v_{nk}-w_{nk}\sum_{t\in T_{nk}}p_{kt}^{*} $\\

(C2) If $u_n^*>0$, then $\sum_{k\in K}x_{nk}^*=1$.\\

The complementary slackness conditions (C1) and (C2) will be (implicitly) used to design the acceptance and rejection rules in the online version of the problem. The online algorithm will maintain {\it integral} primal and dual feasible solutions $\bx$ and $(\bu,\bp)$ such that the dual objective is within a small factor $\alpha$ of the primal objective. It would then follow by {\it weak LP duality} that the \CR\ of the algorithm is at least $\alpha$. Indeed, let $\obj(\cdot)$ be the objective value of the program $(\cdot)$. Then
\begin{align*}
\obj(\textit{IP})\le\obj(P)\le\obj(D)\le\alpha\cdot\obj(P),
\end{align*}
would imply that $\obj(P)\ge\frac{\obj(\textit{IP})}{\alpha}$, which in turn implies, by the fact that the primal solution is integral, that the online algorithm is $\alpha$-competitive.

\subsection{Online Algorithm}
The online algorithm for the basic variant is described in Algorithm \ref{Alg1}. The algorithm constructs simultaneously primal integer and dual solutions. Initially, both primal and dual variables are initialized to zero.  An arriving request $n$ is rejected if $\max_{k \in K} \left\{ v_{nk} - w_{nk} \sum_{t \in T_{nk}} p_{kt} \right\}\le 0$ or assigned to the resource $k^*=\arg\max_{k \in K} \left\{ v_{nk} - w_{nk} \sum_{t \in T_{nk}} p_{kt} \right\}$, where $\bp$ denotes the current price vector (lines~\ref{s1-1}-\ref{s1-5}). If the request is assigned, the dual variables are updated in lines~\ref{s1-6}-\ref{s1-9}.

Recall that, in this basic variant, we use the notation: $\theta_{nk}=\frac{v_{nk}}{w_{nk}d_{nk}}$, $\theta_k^{\max}=\max_{n\in N}\theta_{nk}$, $\theta_k^{\min}=\min_{n\in N:\theta_{nk}>0}\theta_{nk}$, $d_k^{\max}=\max_{n\in N}d_{nk}$, $d_k^{\min}=\min_{n\in N:d_{nk}>0}d_{nk}$, $\bar \theta_k=\frac{\theta_k^{\max}}{\theta_k^{\min}}$, $\bar d_k=\frac{d_k^{\max}}{d_k^{\min}}$, $\bar\theta^{\max}=\max_{k\in K} \bar\theta_k$ and $\bar d^{\max}=\max_{k\in K} \bar d_k$.
In the algorithm, we choose  $\mu_{nk} = e^{\frac{w_{nk}\cdot\gamma_k}{C_k}}$ and $\beta_{nk} = \theta^{\min}_k(\mu_{nk}-1)$, for $n\in N$ and $k\in K$, where
\begin{align}\label{gamma1}
\gamma_k=2 \ln (2 +4\bar\theta_k\bar d_{k}).
\end{align}

\begin{algorithm}[H]
  \caption{Primal-Dual Algorithm for \ORAD}
  \label{Alg1}
  \begin{algorithmic}[1]
    \State Initialize \( x \gets \mathbf{0} \), \( p \gets \mathbf{0} \), \( u \gets \mathbf{0} \)
    \For {each \( n \in N \)}
        \State Compute:        $ k^* \gets \arg\max_{k \in K} \left\{ v_{nk} - w_{nk} \sum_{t \in T_{nk}} p_{kt} \right\} $\label{s1-1}
        \If { \( v_{nk^*} > w_{nk^*} \sum_{t \in T_{nk^*}} p_{k^*t} \) }\label{s1-4}
            \State \( x_{nk^*} \gets 1 \)\label{s1-5}
            \State \( u_n \gets v_{nk^*} - w_{nk^*} \sum_{t \in T_{nk^*}} p_{k^*t} \)\label{s1-6}
            \For{each \( t \in T_{nk^*} \)}
                \State
                \(p_{k^*t} \gets \mu_{nk^*} p_{k^*t} + \beta_{nk^*} \) 
            \EndFor\label{s1-9}
        \EndIf
    \EndFor
  \end{algorithmic}
\end{algorithm}

Note that the acceptance of a request is governed by (C1) and that the primal and dual solutions constructed by the algorithm satisfy (C2). 

\begin{theorem}\label{thm1}
    If $\epsilon_k \leq \frac{C_k \ln 2}{\gamma_k}$ for all $k\in K$, Algorithm~\ref{Alg1} outputs a feasible solution and has a competitive ratio of $ O\big(\log (\bar\theta^{\max}\cdot\bar d^{\max})\big)$. 
\end{theorem}

We prove theorem \ref{thm1} by showing that the primal and dual solutions obtained by Algorithm~\ref{Alg1} satisfy the following lemmas. 

\begin{lemma} \label{Lemma1} 
(a) The dual solution generated by Algorithm \ref{Alg1} is feasible.   \\
(b) For all $k\in K$ and $t\in T$, 
   $p_{kt} = \theta^{\min}_k\left(e^{\frac{\gamma_{k} \cdot z_{kt}}{C_{k}}} - 1\right)$, 
where $z_{kt}= \sum_{n\in N:t\in T_{nk}} w_{nk}x_{nk}$ is the {\it utilization} at time slot $t$ of resource $k$.

\end{lemma}

Let  $\Delta P$ and $\Delta D$ be the changes in the primal and dual objective values in one iteration.

\begin{lemma}
(a) If an arriving request is not assigned to a resource, there is no change in the primal or the dual solution.\\
(b) If an arriving request is assigned to a resource $k^*$, $\frac{\Delta D}{\Delta P}$ is bounded by 
 $\frac{2\gamma_{k^*}}{ \ln 2}$ .\label{Lemma2}
\end{lemma}

\begin{lemma}
The primal solution generated by  Algorithm \ref{Alg1} is feasible.\label{Lemma3}
\end{lemma}

\sproofof{Lemma~\ref{Lemma1}}
(a) Consider request $n$ and $k^*$ defined in line \ref{s1-1} of Algorithm  \ref{Alg1}. Upon the arrival of the request, $u_n=0$ if $v_{nk^*}-w_{nk^*}\sum_{t\in T_{nk^*}}p_{k^*t}\leq 0$,  and  $u_n=v_{nk^*}-w_{nk^*}\sum_{t\in T_{nk^*}}p_{k^*t}$ otherwise. For any other resource $k\in K$, we have constraint (\ref{constr_dual: u_n}) satisfied by the choice of $k^*$. Since the value of $u_n$ does not change in subsequent steps, $u_n\geq 0$ is maintained. As further increases of $p_{kt}$ cannot lead to infeasibility, constraint (\ref{constr_dual: u_n}) remains satisfied throughout the algorithm. Furthermore, as $\mu_{nk}\geq 0$ and $ \beta_{nk}\geq 0$, $p_{kt}\geq 0$ throughout the algorithm.  

(b) We prove the statement by induction on the iterations of Algorithm~\ref{Alg1}.  Initially, when no request is occupying  time slot $t$ in resource $k$ then both $z_{kt} = 0$ and  $p_{kt} = 0$. 

Consider now a request $n$ that is admitted to occupy slot $t$ of resource $k$. Denote by $\bz$, $\bp$ and $\bar{\bz}$, $\bar{\bp}$, the utilization and price vectors right before and after request $n$ was admitted, respectively. By the update policy of Algorithm~\ref{Alg1}, 
\begin{equation}
   \bar p_{kt} = \mu_{nk} p_{kt} + \beta_{nk}.\label{update}
\end{equation} 
Then $\bar z_{kt}=z_{kt}+w_{nk}$ for all $n$ and $k$.
Combining the induction hypothesis and  (\ref{update}), we obtain
\begin{align*}
      \bar p_{kt} &= \mu_{nk} \theta^{\min}_k(e^{\frac{\gamma_k\cdot z_{kt}}{C_{k}}} - 1) + \beta_{nk}=\mu_{nk} \theta^{\min}_k(e^{\frac{\gamma_k\cdot z_{kt}}{C_{k}}} - 1) + \theta^{\min}_k(\mu_{nk}-1)\\
      &= \theta^{\min}_ke^{\frac{w_{nk}\cdot\gamma_{k}}{C_{k}}} \left(e^{\frac{\gamma_{k}\cdot z_{kt}}{C_{k}}} - 1 \right) + \theta^{\min}_ke^{\frac{w_{nk}\cdot\gamma_{k}}{C_{k}}} - \theta^{\min}_k
      =\theta^{\min}_k\left(e^{\frac{\gamma_{k} \cdot\bar{z}_{kt}}{C_{k}}} - 1\right),
\end{align*}
finalizing the induction proof.
\qed

\medskip

\sproofof{Lemma~\ref{Lemma2}}
 (a)  If in an iteration, the arriving request is not assigned to a resource, $\Delta D=\Delta P=0$. \\
(b) Assume that request $n$ is assigned to a resource $k^*$. The change in the primal objective value is 
$\Delta P = v_{nk^*}$, 
while the change in the dual objective value is 
\begin{equation}
    \Delta D 
 =u_n +  C_{k^*}\sum_{t\in T_{nk^*}} [\mu_{nk^*} p_{k^*t}+ \beta_{nk^*} - p_{k^*t}].
 \label{dual_change1}    
\end{equation}
Substituting  $u_n = v_{nk^*} - w_{nk^*}\sum_{t \in T_{nk^*}}  p_{k^*t}$, and taking into account that $|T_{nk^*}|= d_{nk^*}$, (\ref{dual_change1}) becomes 
\begin{equation}
    \Delta D = v_{nk^*} + [C_{k^*}(\mu_{nk^*}-1) - w_{nk^*}]\sum_{t \in T_{nk^*}} p_{k^*t} + C_{k^*} \beta_{nk^*} d_{nk^*}.\label{dual_change2}
\end{equation}
Recall that $\mu_{nk^*} = e^{\frac{w_{nk^*}\cdot\gamma_{k^*}}{C_{k^*}}}$ and $\beta_{nk^*} = \theta_{k^*}^{\min}(\mu_{nk^*}-1)$. The choice~\raf{gamma1} implies that $\gamma_{k^*}\geq 1$ and therefore $\mu_{nk^*}\geq 1+\frac{w_{nk^*}}{C_{k^*}}$. By step~\ref{s1-4} of the algorithm, $\sum_{t\in T_{nk^*}} p_{k^*t} < \frac{v_{nk^*}}{w_{nk^*}}$. Finally, from the definition of $\theta_{k^*}^{\min}$, $d_{nk^*} \leq \frac{v_{nk^*}}{w_{nk^*}\theta_{k^*}^{\min}}$. Using these bounds in (\ref{dual_change2}), we obtain

\begin{equation}
    \Delta D \leq  \frac{2C_{k^*}(e^{\frac{w_{nk^*}\cdot\gamma_{k^*}}{C_{k^*}}}-1)v_{nk^*}}{w_{nk^*}}\label{delta_d_}.
\end{equation}

From the bounds on $\Delta P$ and $\Delta D$, it follows that 
\begin{equation}
    \frac{\Delta D}{\Delta P} \leq \frac{2C_{k^*}(e^{\frac{w_{nk^*}\cdot\gamma_k}{C_{k^*}}}-1)}{w_{nk^*}}.\label{D_P_ratio}
\end{equation}
Note that by (A3) and our assumption on $\epsilon_{k^*}$, $\frac{w_{nk^*}\cdot\gamma_{k^*}}{C_{k^*} \ln 2}\in [0,1]$. As for $x\in [0, 1]$, $e^{x \ln 2}-1 \leq x$  we obtain    
\begin{equation}
    e^{\frac{w_{nk^*}\cdot\gamma_{k^*}}{C_{k^*}}}-1 \leq \frac{w_{nk^*}\cdot\gamma_{k^*}}{C_{k^*} \ln 2}.\label{exp}
\end{equation}
Combining (\ref{D_P_ratio})  and  (\ref{exp}) we obtain $
    \frac{\Delta D}{\Delta P} \leq \frac{2\gamma_{k^*}}{ \ln 2}.$
The lemma follows.
\qed

\medskip

\sproofof{Lemma~\ref{Lemma3}}
 To prove that the primal solution obtained by Algorithm \ref{Alg1} is feasible, we need to show that the algorithm stops assigning requests to a resource once that resource is full. For the sake of contradiction, suppose that there exists a $t_0\in T$ and $k^*\in K$ such that resource $k^*$ becomes overloaded at time slot $t_0$, i.e., $\sum_{n\in N: t_0 \in T_{nk^*}}w_{nk^*}x_{nk^*} > C_{k^*}$. We will show that this implies that there is a request $n_0\in N$ that got assigned to $k^*$ such that $w_{n_0k^*}\sum_{t \in T_{n_0k^*}} p_{k^*t} > v_{n_0k^*}$, where $p_{k^*t}$ is the price vector right before admitting $n_0$. This contradicts to our admission policy in line~\ref{s1-4} of the algorithm. 

 Let $I$ be the set of accepted requests that occupy slot $t_0$ of resource $k^*$, i.e. $I = \{n: x_{nk^*} = 1,~t_0\in T_{nk^*}\}$. Define $
    I_1 = \left\{n \in I: t_0- s_{nk^*} \geq \frac{d_{nk^*}}{2}\right\} $ (that is, the set of requests that have most of their duration on the left of $t_0$)
 and $I_2=I\setminus I_1$.
Without loss of generality, we assume that $  \sum_{n \in I_1} w_{nk^*}\geq \sum_{n \in I_2} w_{nk^*}.
$
\begin{figure}[H]
	\centering
\includegraphics[width=1\textwidth]{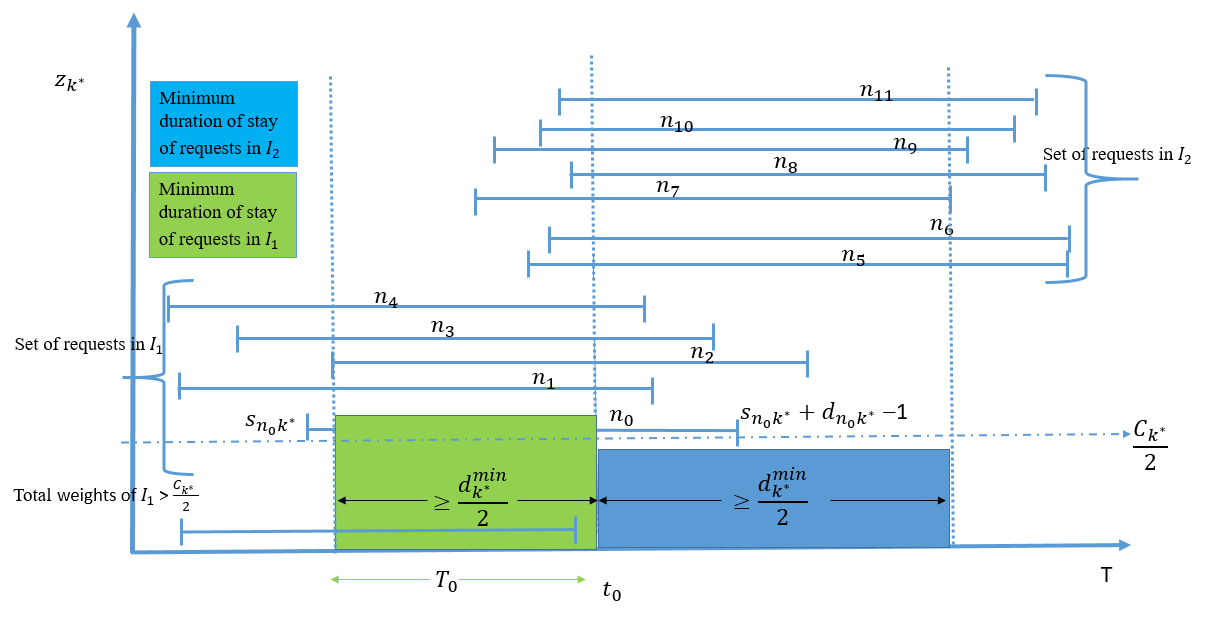}
	\caption{Illustration of the worst-case of final capacity utilization.}
	\label{fig_}
\end{figure}
Note that by construction of $I_1$, it is easy to see that all requests in $I_1$ occupy the resource $k^*$ for a duration of at least $\frac{d^{min}_{k^*}}{2}$  successive slots.  See Figure \ref{fig_} for an illustration.
Since at $t_0$, $\sum_{n\in N: t_0 \in T_{nk^*}}w_{nk^*}x_{nk^*} > C_{k^*}$, we get $
   \sum_{n \in I_1} w_{nk^*}  > \frac{C_{k^*}}{2}.$
Recall that $z_{k^*t}= \sum_{n:t\in T_{nk^*}} w_{nk^*}x_{nk^*}$. Let $n_0$ be the last admitted request in $I_1$ with strictly positive weight, i.e., $w_{n_0k^*}>0$. Right before admitting $n_0$, we have
\begin{equation}
    z_{k^*t} \geq \sum_{n \in I_1 \setminus \{n_0\}} w_{n k^*}, \quad \forall~ t \in T_0 = \bigcap\limits_{n \in I_1} T_{n k^*}.\label{z_lowerbound}
\end{equation}
Combining with (A3) we obtain,
\begin{equation}
  z_{k^*t} > \frac{C_{k^*}}{2}- w_{n_0k^*} \geq \frac{C_{k^*}}{2} - \epsilon_{k^*} \geq \frac{C_{k^*}}{2} - \frac{C_{k^*} \ln 2}{\gamma_{k^*}}. \label{util_simp} 
\end{equation}


By Lemma~\ref{Lemma1}(b), the price right before admitting $n_0$ satisfies $p_{k^*t} = \theta^{\min}_{k^*}\left(e^{\frac{\gamma_{k^*} \cdot z_{k^*t}}{C_{k^*}}} - 1 \right)$ for any  $t \in T_0$. Together with (\ref{util_simp}) this implies
\begin{align*}
   \sum_{t \in T_{n_0k^*}} p_{k^*t} &= \theta^{\min}_{k^*}\sum_{t \in T_{n_0k^*}} \left( e^{\frac{\gamma_{k^*}\cdot z_{k^*t}}{C_{k^*}}} - 1  \right) \geq \theta^{\min}_{k^*}\sum_{t \in T_0} \left( e^{\frac{\gamma_{k^*}\cdot z_{k^*t}}{C_{k^*}}} - 1  \right)\\
   &> \theta^{\min}_{k^*}\sum_{t \in T_0}  \left( e^{\frac{\gamma_{k^*} \left(\frac{C_{k^*}}{2} - \frac{C_{k^*} \ln 2}{\gamma_{k^*}}\right)}{C_{k^*}}} - 1  \right)= \theta^{\min}_{k^*}|T_0 | \left( e^{\left(\frac{\gamma_{k^*} }{2} -  \ln 2\right)} - 1  \right).\label{sum_p}
\end{align*}

By taking into account that $|T_0| \geq \frac{d^{min}_{k^*}}{2}$, we obtain
\begin{equation}
 \sum_{t \in T_{n_0k^*}} p_{k^*t} >  \frac{\theta^{\min}_{k^*}d^{min}_{k^*}}{2} \left( e^{\left(\frac{\gamma_{k^*} }{2} -  \ln 2\right)} - 1  \right).\label{sum_p_1}
\end{equation}

By definition, $\gamma_{k^*} = 2 \ln (2 +4\bar\theta_{k^*}\bar d_{k^*})$ where $\bar\theta_{k^*} =\frac{\theta_{k^*}^{\max}}{\theta_{k^*}^{\min}}$ and $\bar d_{k^*}=\frac{d^{max}_{k^*}}{d^{min}_{k^*}}$. Hence, 
\begin{align}
\gamma_{k^*} &=2 \ln \left(2 +4\cdot\frac{\theta_{k^*}^{\max}}{\theta^{\min}_{k^*}}\cdot\frac{d^{max}_{k^*}}{d^{min}_{k^*}} \right)
\ge2\ln\left(2+4\cdot\frac{v_{n_0k^*}}{w_{n_0k^*}}\cdot\frac1{\theta^{\min}_{k^*}\cdot d^{\min}_{k^*}}\right)
 \label{lower_bound_gamma}.
\end{align}
Combining (\ref{sum_p_1}) and (\ref{lower_bound_gamma}) we obtain
\begin{equation}\label{e4}
 \sum_{t \in T_{n_0k^*}} p_{k^*t} > \frac{\theta^{\min}_{k^*}d^{min}_{k^*}}{2} \left( \frac{v_{n_0k^*}}{w_{n_0k^*}}\cdot\frac1{\theta^{\min}_{k^*}\cdot d^{\min}_{k^*}} \right) = \frac{v_{n_0k^*}}{w_{n_0k^*}},
\end{equation}
contradicting the assumption that $n_0$ was admitted by our algorithm. We conclude that the primal solution of Algorithm \ref{Alg1} remains feasible throughout the algorithm. 
\qed

\medskip

\sproofof{Theorem \ref{thm1}}
It follows from the proof of the Lemmas \ref{Lemma1}-\ref{Lemma3} that the competitive ratio (\CR) of Algorithm \ref{Alg1} is 
$
  \CR = \frac{\Delta D}{\Delta P} = O\big(\log (\bar\theta^{\max}\cdot\bar d^{\max})\big).$
Also, the algorithm outputs an integral solution which is is feasible by Lemma~\ref{Lemma3}.
\qed
\section{The Load Balancing Variant}\label{sec:load}
As in the standard variant, each request $n\in N$ is defined by a tuple $(a_n,v_{nk},w_{nk},s_{nk},d_{nk})$. Additionally, requests arrive in batches. Let $N_i=\{n\in N:~a_n=i\}$ be the batch of requests arriving at time $i$ and  $q_k\in\ZZ_+$ be given numbers for $k\in K$. As a form of balancing the load among resources,  when batch $N_i$ arrives (at time $i$), it is constrained that no more than $q_k$ requests can be allocated to resource $k$. 

The primal and dual LPs corresponding to the integer programming formulation are given by



{\centering \hspace*{-10pt}
\scriptsize
\begin{minipage}[t]{.55\textwidth}
  \begin{alignat}{2}
    (P^{\text{lb}}) \quad & \max_{\mathbf{x}}  \sum_{n \in N,k \in K} v_{nk} x_{nk}\nonumber\\
    \text{s.t.} \quad 
    & \sum_{n \in N : t \in T_{nk}} w_{nk} x_{nk}  \leq C_k \quad \forall~ k \in K, \forall~ t \in T_{nk}, \nonumber
    \\
    & \sum_{k \in K} x_{nk} \leq 1 \quad \forall~ n \in N, \nonumber
    \\
    & \sum_{n \in N_i} x_{nk}  \leq q_k \quad \forall~ k \in K, \forall~ i \in T, \label{eq:assignment_rel2} \\
    &x_{nk}  \geq 0 \quad \forall~ n \in N, k \in K.\nonumber
  \end{alignat}
\end{minipage}
\hspace{0.3em}
\vrule width 0.7pt
\hspace*{-5pt}
\hspace{0.3em}
\begin{minipage}[t]{.55\textwidth}
  \begin{alignat}{2}
    (D^{\text{lb}}) \quad & \min_{\mathbf{u}, \mathbf{p}, \mathbf{h}}  \sum_{k \in K,t \in T_{nk}} C_k p_{kt} + \sum_{n \in N} u_n + \sum_{k \in K,i \in T} q_k h_{ki}\nonumber\\
    \text{s.t.} \quad 
    & w_{nk} \sum_{t \in T_{nk}} p_{kt} + u_n + h_{ki}  \geq v_{nk} \quad \forall~ n \in N, k \in K, \nonumber\\
    & u_n  \geq 0 \quad \forall~ n \in N, \nonumber\\
    & p_{kt}\geq 0 \quad \forall~ k \in K, t \in T_{nk},\nonumber\\
    &h_{ki}  \geq 0 \quad \forall~i \in T\nonumber.
  \end{alignat}
\end{minipage}}
\vspace{.1in}

  The additional constraints (\ref{eq:assignment_rel2}) ensure that for each resource $k$, the total number of requests accepted at time $i$ from the batch $N_i$ does not exceed  $q_k$. 
 The  variables 
 $h_{k,i}$ are the dual variables corresponding to constraints 
 (\ref{eq:assignment_rel2}).

The primal-dual algorithm given in Algorithm \ref{Alg_LB} proceeds as follows. In line~\ref{s2-1} of the algorithm, both primal and dual variables are initialized to $\bzero$. On arrival of a new batch of requests $N_i$ at time $i$,  the algorithm 
solves the linear program ($P_i$) below, which is guaranteed to have an integer optimal solution (as the constraints matrix is {\it totally unimodular}).  Let $\mathbf{\bar{x}}$ be the optimal solution of ($P_i$).  If $\bar{x}_{nk}=1$, assign request $n$ to resource $k$ and update the dual solution in lines \ref{algloadupdatedual:start}-\ref{algloadupdatedual:end}. 
When updating the dual variables $p_{kt}$,  we choose  $\mu_{nk} = e^{\frac{w_{nk}\cdot\gamma_k}{C_k}}$ and $\beta_{nk} = \theta_k^{\min}(\mu_{nk}-1)$, for $n\in N$ and $k\in K$, where $\gamma_k$ is set as in~\raf{gamma1}.

\vspace{-.1in}
{\centering \hspace*{-18pt}
\scriptsize
\begin{minipage}[t]{.45\textwidth}
  \begin{alignat}{3}
	(P_i) \quad& \displaystyle \max_{\bx} \quad  \sum_{n \in N_i,k \in K:~r_{nk}>0} r_{nk} x_{nk} \nonumber\\
		\text{s.t.} 
\quad & \sum_{n\in N_i} x_{nk} \leq q_k \quad \forall~ k \in K, 
        \nonumber \\
&\sum_{k \in K} x_{nk} \leq 1 \quad \forall~ n \in N_i, 
        \nonumber \\
& x_{nk}\ge 0\quad \forall~ n \in N_i,~ k \in K, 
\nonumber
   \end{alignat}
   \end{minipage}
	\,\,\, \rule[-28ex]{1pt}{28ex}
     \hspace{0in}
	\begin{minipage}[t]{0.45\textwidth}
		\begin{alignat}{3}
		(D_i) \quad &\displaystyle \min_{\bu,\bp,\bh}\quad     
      \sum_{n\in N_i}u_{n}+\sum_{k\in K}q_kh_{k,i}  \nonumber\\
		\text{s.t} \quad & u_n + h_{ki}\geq r_{nk}\quad \forall~ n\in N_i,~k\in K\text{ s.t. } r_{nk}>0,\nonumber 
 \\
 &u_{n} \geq 0 \quad \forall~ n\in N_i,\nonumber\\
 &h_{ki} \geq 0 \quad \forall~ k\in K. \nonumber
    \end{alignat}
\end{minipage}}

\medskip

Note that ($P_i$) includes only pairs $n\in N_i$ and $k\in K$ for which $r_{nk}>0$, where $r_{nk}=v_{nk} - w_{nk} \sum_{t \in T_{nk}} p_{kt}$. In particular, pairs $(n,k)$ with $r_{nk}\le0$ never get assigned.
\begin{algorithm}[H]
  \caption{Primal-Dual Algorithm for \lbORAD}
  \label{Alg_LB}
  \begin{algorithmic}[1]
    \State Initialize \( \bx \gets \mathbf{0} \), \( \bp \gets \mathbf{0} \), \( \bu \gets \mathbf{0} \), \( \bh \gets \mathbf{0} \) \label{s2-1}   
    \For {each time slot \( i \in T \)}
        \For {each request \( n \in N_i \)}
            \For {each resource \( k \in K \)}
                \State $r_{nk} \gets v_{nk} - w_{nk} \sum_{t \in T_{nk}} p_{kt}$
            \label{s2-5}
            \EndFor
        \EndFor


        \State Find an optimal integer solution  $\bar{\bx}$ for ($P_i$) and an optimal solution $(\bar{\bu}, \bar{\bh})$ for   ($D_i$)

        \For {each \( n \in N_i \)} \label{algloadupdatedual:start}
            \State Let $u_n=\bar{u}_n$
            \For {each \( k \in K \)} 
            \State  $h_{ki}\gets\bar{h}_{ki}$
                \If { \( \bar{x}_{nk} = 1 \) }
                \State  $x_{nk}\gets 1$
                    \For {each \( t \in T_{nk} \)}
                        \State \( p_{kt} \gets \alpha_{nk}  p_{kt} + \beta_{nk} \)
                    \EndFor
                \EndIf
            \EndFor
        \EndFor \label{algloadupdatedual:end}
    \EndFor
  \end{algorithmic}
\end{algorithm}

\begin{theorem}\label{thm2}
If $\epsilon_k \leq \frac{C_k \ln 2}{q_k\gamma_k}$ for all $k\in K$, Algorithm \ref{Alg_LB} outputs a feasible solution and has a competitive ratio of $ O\big(\log (\bar\theta^{\max}\cdot\bar d^{\max})\big)$.
\end{theorem}
To prove Theorem \ref{thm2}, we show that the primal and dual solutions obtained by Algorithm \ref{Alg_LB} satisfy the following lemmas.

\begin{lemma} \label{Lemma4} 
(a) The dual solution generated by Algorithm \ref{Alg_LB} is feasible.

(b) For $k\in K$ and $t\in T$, $p_{kt} = \theta_k^{\min}\left(e^{\frac{\gamma_{k} \cdot z_{kt}}{C_{k}}} - 1\right)$, 
where  $z_{kt} = \sum_{n\in N: t\in T_{nk}}w_{nk}x_{nk}$.  
\end{lemma}
Let  $\Delta P^{\text{ld}}$ and $\Delta D^{\text{ld}}$ be the changes in the primal and dual objective values in one iteration.
\begin{lemma}
(a) If in iteration $i$, no request gets assigned to a resource, there is no change in the primal or the dual solution.\\
(b) If in iteration $i$, some requests get assigned to some resources, $\frac{\Delta D^{\text{ld}}}{\Delta P^{\text{ld}}}$ is bounded by $\frac{2\gamma^{\max}}{ \ln 2}$, where $\gamma^{max}=\max_{k\in K}\gamma_{k}$.
\label{Lemma5}
\end{lemma}

\begin{lemma}
The primal solution generated by  Algorithm \ref{Alg_LB} is feasible.\label{Lemma6}
\end{lemma}

\medskip

\sproofof{Lemma~\ref{Lemma4}} 
The proof is similar to that of Lemma~\ref{Lemma1} and is omitted.
\qed

\medskip

\sproofof{Lemma~\ref{Lemma5}}
(a)  At time $i$, consider a request $n\in N_i$ and a resource $k\in K$. From the feasibility of $\bar{\bu}, \bar{\bh}$ for ($D_i$) and the definition of $r_{nk}$, $u_n$ and $h_{ki}$, we obtain  $u_n+h_{ki}=\bar{u}_n + \bar{h}_{ki}\geq r_{nk}= v_{nk} - w_{nk}\sum_{t \in T_{nk}}  p_{kt}$, if $r_{nk}>0$. This implies that  $w_{nk}\sum_{t\in T_{nk}}p_{kt}+u_n + h_{ki}\geq v_{nk}$ for all $n$, $k$ and $i$. As further increase of the price values $p_{kt}$ can not lead to infeasibility, we conclude that the dual solution obtained is feasible.

(b) can be proven by induction, similar to Lemma \ref{Lemma1}(b).
\qed

\medskip

\sproofof{Lemma~\ref{Lemma5}}
Consider iteration $i$ of Algorithm \ref{Alg_LB}. The change in the primal objective value is $\Delta P^{\text{ld}}= \sum_{n\in N_i} v_{nk}x_{nk}$, 
while the change in dual objective value is 
\begin{align}
    \Delta D^{\text{ld}} &= \sum_{n\in N_i}u_{n} +\sum_{k\in K}q_kh_{ki} + \sum_{n \in N_i,k \in K}C_k\left[ ( \mu_{nk} -1)\sum_{t\in T_{nk}}p_{kt}+ \sum_{t\in T_{nk}}\beta_{nk} \right]x_{nk}\nonumber \\
    &= \sum_{n\in N_i}u_{n} +\sum_{k\in K}q_kh_{ki} + \sum_{n \in N_i,k \in K}C_k\left[ ( \mu_{nk} -1)\sum_{t\in T_{nk}}p_{kt}+ \beta_{nk}d_{nk} \right]x_{nk}.\label{delta_D1}
\end{align}
By strong duality for the primal-dual pair of LPs ($P_i,D_i$),
\begin{equation*}
     \sum_{n \in N_i,k \in K:~r_{nk}>0} r_{nk} \bar{x}_{nk} =  \sum_{n\in N_i}\bar{u}_{n}+\sum_{k\in K}q_k\bar{h}_{ki}.
    \end{equation*}
As $\bar{x}_{nk}={x}_{nk}$ and  $u_n=\bar{u}_n$, for $n\in N_i$,  and  $h_{ki}=\bar{h}_{ki}$, for $k\in K$ and $i\in T$, (\ref{delta_D1}) becomes
\begin{equation}
    \Delta D^{\text{ld}} = \sum_{n \in N_i,k \in K:~r_{nk}>0} r_{nk} x_{nk} + \sum_{n \in N_i,k \in K}C_k\left[ ( \mu_{nk} -1)\sum_{t\in T_{nk}}p_{kt}+ \beta_{nk}d_{nk} \right]x_{nk}.
\end{equation}
Recall from Algorithm~\ref{Alg_LB} that $x_{nk}$ can be positive only if $r_{nk}>0$, where $r_{nk} = v_{nk} - w_{nk} \sum_{t \in T_{nk}} p_{kt}$. Hence, 
\begin{align}
    \Delta D^{\text{ld}} &= \sum_{n \in N_i,~k \in K} \left(v_{nk} - w_{nk} \sum_{t \in T_{nk}} p_{kt} \right)x_{nk} + \sum_{n \in N_i,k \in K}C_k\left[ ( \mu_{nk} -1)\sum_{t\in T_{nk}}p_{kt}+ \beta_{nk}d_{nk} \right]x_{nk}\nonumber\\
    &= \sum_{n \in N_i,k \in K}v_{nk}x_{nk} + \sum_{n \in N_i,k \in K} \left[ [C_k(\mu_{nk} -1) - w_{nk}]x_{nk}\cdot\sum_{t\in T_{nk}}p_{kt} +C_k \beta_{nk}d_{nk}x_{nk}  \right].\nonumber
\end{align}
As $x_{nk}=1$ implies $r_{nk}>0$ and hence $v_{nk} - w_{nk} \sum_{t \in T_{nk}} p_{kt} > 0$, we have $\sum_{t \in T_{nk}} p_{kt} \leq \frac{v_{nk}}{w_{nk}}$ if $x_{nk}=1$, and thus,
\begin{align}
    \Delta D^{\text{ld}}  &\leq \sum_{n \in N_i,k \in K}v_{nk}x_{nk} + \sum_{n \in N_i,k \in K} \left[ [C_k(\mu_{nk} -1) - w_{nk}]x_{nk}\cdot\frac{v_{nk}}{w_{nk}} +C_k \beta_{nk}d_{nk}x_{nk}  \right]\nonumber\\
    & \leq \sum_{n \in N_i,k \in K}v_{nk}x_{nk} + \sum_{n \in N_i,~k \in K} \left[ [C_k(\mu_{nk} -1) - w_{nk}]x_{nk}\cdot\frac{v_{nk}}{w_{nk}} +C_k \theta_k^{\min}(\mu_{nk} -1)d_{nk}x_{nk}  \right]\nonumber\\
    & = \sum_{n \in N_i,k \in K} C_k(\mu_{nk} -1) \left[\frac{v_{nk}}{w_{nk}} + \theta_k^{\min}d_{nk} \right]x_{nk}. \label{eqn_46}
\end{align}
Recall that $\mu_{nk} = e^{\frac{w_{nk}\cdot\gamma_k}{C_k}}$. As in (\ref{exp}), $\mu_{nk}-1 \leq  \frac{w_{nk}\cdot\gamma_{k}}{C_{k} \ln 2}$. Also, by assumption (A3), $\left[\frac{v_{nk}}{w_{nk}} + \theta_k^{\min}d_{nk} \right] \leq \frac{2v_{nk}}{w_{nk}}$. Thus, (\ref{eqn_46}) gives
\begin{equation*}\label{e3}
   \Delta D^{\text{ld}} \leq  \sum_{n \in N_i,k \in K}  \frac{2\gamma_k }{ \ln 2}\cdot v_{nk}x_{nk}\leq \frac{ 2\gamma^{max}}{\ln 2}\sum_{n \in N_i,k \in K}v_{nk}x_{nk}.  
\end{equation*}
This yields $\frac{\Delta D^{\text{ld}}}{\Delta P^{\text{ld}}}\leq \frac{2\gamma^{max}}{\ln 2}.$
\qed



\medskip

\sproofof{Lemma~\ref{Lemma6}}
 To prove that the primal solution of Algorithm \ref{Alg_LB} is feasible, we need to show that whenever the capacity of the resource is full, then the algorithm stops assigning batches of requests to that resource. For the sake of contradiction, suppose that there is  $t_0\in T$ and $k\in K$ such that a violation happens at time slot $t_0$ for resource $k$, i.e., we have $\sum_{n\in N: t_0 \in T_{nk}}w_{nk}x_{nk} > C_{k}$. We will show that there exists a time instance $i$ and an accepted batch of requests $N' \subset N_i$, all assigned to resource $k$, such that $w_{nk}\sum_{t \in T_{nk}} p_{kt} > v_{nk}$, that is, $r_{nk}<0$, for all $n\in N'$, where $p_{kt}$ is the price vector right before admitting $N'$, in contradiction to our assignment policy which only assigns requests with $r_{nk}>0$. \\
     
 Let $I$ be the set of accepted requests that occupy slot $t_0$ of resource $k$, i.e., $I = \{n: x_{nk} = 1,~t_0\in T_{nk}\}$. Define $
    I_1 = \left\{n \in I: t_0- s_{nk} \geq \frac{d_{nk}}{2}\right\} $
 and  $I_2=I\setminus I_1$.
   Without loss of generality, suppose that 
\begin{equation*}
    \sum_{n\in I_1} w_{nk} > \sum_{n\in I_2} w_{nk},
\end{equation*}
which implies that $\sum_{n\in I_1} w_{nk}> \frac{C_k}{2}$.

Let us consider the last batch $N' \subseteq  N_i\cap I_1$  that was admitted to resource $k$ (at time $i$), where $|N'|\leq q_k$.

Hence, prior to admitting $N'$, we have that 
\begin{equation}
    z_{kt} \geq \sum_{n\in I_1 \setminus N'}w_{nk}> \frac{C_k}{2} - |N'|\epsilon_k, \forall t\in  T_0 = \bigcap\limits_{n \in I_1} T_{n k},\label{lowerboundz}
\end{equation}
   where  $z_{kt} = \sum_{n: t\in T_{nk}}w_{nk}x_{nk}$ is the utilization prior to admitting $N'$.
By Lemma \ref{Lemma4}(b), the price right before admitting $N'$ satisfies 
\begin{equation}
        p_{kt} = \theta_k^{\min}\left( {e^{\frac{\gamma_{k}\cdot z_{kt}}{C_{k}}}} - 1\right).\label{p_feasible1}
    \end{equation}
By summing (\ref{p_feasible1}) for any $n\in N'$ over $t \in T_{nk}$, and using (\ref{lowerboundz}), we obtain
\begin{align}
   \sum_{t \in T_{nk}} p_{kt} &= \theta_k^{\min}\sum_{t \in T_{nk}} \left( e^{\frac{\gamma_{k}\cdot z_{kt}}{C_{k}}} - 1  \right) \geq \theta_k^{\min}\sum_{t \in T_0} \left( e^{\frac{\gamma_{k}\cdot z_{kt}}{C_{k}}} - 1  \right)\nonumber\\
   &> \theta_k^{\min}\sum_{t \in T_0}  \left( e^{\frac{\gamma_{k} \left(\frac{C_k}{2} - |N'|\epsilon_k\right)}{C_{k}}} - 1  \right)\nonumber\\
   &= \theta_k^{\min}|T_0| \left( e^{\gamma_k  \left(\frac{1}{2}-\frac{|N'|\epsilon_k}{C_k} \right)} - 1  \right).\label{sum_p1}
\end{align}
By taking into account that $|T'| \geq \frac{d^{\min}_{k}}{2}$ in
(\ref{sum_p1}), we obtain
\begin{equation}
 \sum_{t \in T_{nk}} p_{kt} > \frac{\theta_k^{\min}\cdot d^{min}_{k}}{2} \left( e^{\gamma_k  \left(\frac{1}{2}-\frac{|N'|\epsilon_k}{C_k} \right)} - 1  \right)\geq \frac{\theta_k^{\min}\cdot d^{min}_{k}}{2} \left( e^{\gamma_k  \left(\frac{1}{2}-\frac{q_k\cdot\epsilon_k}{C_k} \right)} - 1  \right). \label{sum_p_12}
\end{equation}

Recall that $\gamma_k= 2\ln (2 + 4\bar\theta^{\max}\cdot\bar d^{\max})$ and that $\epsilon_k\leq \frac{C_k\ln 2}{q_k\cdot\gamma_k}$ by our assumption.
Thus, (\ref{sum_p_12}) yields 
\begin{align}
 \sum_{t \in T_{nk}} p_{kt} &>  \frac{\theta_k^{\min} \cdot d^{min}_{k}}{2} \left( e^{\gamma_k  \left(\frac{1}{2}-\frac{q_k\epsilon_k}{C_k} \right)} - 1  \right) 
 \geq \frac{\theta^{\min}_{k^*}d^{min}_{k^*}}{2} \left( e^{\left(\frac{\gamma_{k} }{2} -  \ln 2\right)} - 1  \right).\label{sum_p_21}
\end{align}
As in~\raf{lower_bound_gamma}, we obtain by the choice of $\gamma_k$ that
\begin{equation*}
 \sum_{t \in T_{nk}} p_{kt} > \frac{v_{nk}}{w_{nk}},
\end{equation*}
contradicting the assumption that $n\in N'$ was admitted by our algorithm. We conclude that the primal solution of Algorithm \ref{Alg_LB} remains feasible throughout the algorithm. 
\qed

\medskip

\sproofof{Theorem \ref{thm2}}
It follows from Lemmas~\ref{Lemma4}-\ref{Lemma6} that the competitive ratio (\CR) of Algorithm \ref{Alg1} is 
$
  \CR = \frac{\Delta D^{\text{ld}}}{\Delta P^{\text{ld}}}= O\big(\log (\bar\theta^{\max}\cdot\bar d^{\max})\big).$ Also, the algorithm outputs an integral solution which is feasible by Lemma~\ref{Lemma6}.

\section{The Multi-dimensional Multiple Knapsack Variant}\label{sec:multi}


In this variant, we consider the problem of allocating a set of requests $N$ to a set of multi-dimensional resources $K$.
When a request $n \in N$ is admitted, it is assigned to a resource $k \in K$ and consumes multiple resource types (or dimensions), indexed by $m \in M_k$, each specified by its time interval $T_{nkm}$ of length $d_{nmk}$. Thus, the resource consumption is specified by a weight vector $\bw_{nk} = (w_{nkm})_{m \in M_k}$ and an interval vector $\bT_{nk}=(T_{nkm})_{m
\in M_k}$, and each resource $k\in K$ has a capacity vector $\mathbf{C}_{k} = \{ C_{km} \}_{m \in M_k}$, where $M_k$ denotes the set of resource dimensions associated with resource $k$. Upon acceptance of a request $n$ to a resource $k$, a reward of $v_{nk}$ is obtained. The corresponding primal and dual linear programs derived from the underlying integer program are given by:


{\centering 
\begin{minipage}[t]{.45\textwidth}
  \begin{alignat}{3}
	(P^{\text{mmd}}) \quad& \displaystyle \max_{\mathbf{x}} \quad  \sum_{n \in N,k \in K} v_{nk} x_{nk} \nonumber\\
	\text{s.t.} \quad & 
\sum_{n \in N: t \in T_{nkm}} w_{nkm} x_{nk} \leq C_{km} \quad \forall~ k \in K,~ m\in M_k,~ \forall~ t \in T \nonumber,\\
&\sum_{k\in K}x_{nk}\leq 1 \quad \forall n\in N ,\nonumber\\
& x_{nk}\ge 0\quad \forall~ n \in N,~ k\in K. \nonumber
   \end{alignat}
\end{minipage}\\
\begin{minipage}[t]{0.45\textwidth}
  \begin{alignat}{3}
  (D^{\text{mmd}}) \quad& \min_{\bu,\bp}\quad \sum_{k\in K,m\in M_k,t\in T} C_{km} p_{kmt} + \sum_{n\in N} u_{n} \nonumber\\
  \text{s.t.} \quad & \sum_{m\in M_k,t\in T_{nkm}} w_{nkm} p_{kmt} + u_n \geq v_{nk} \quad \forall~ n\in N,~k\in K, \nonumber \\
  & u_{n} \geq 0 \quad \forall~ n\in N, \nonumber\\
  & p_{kmt} \geq 0 \quad \forall~ k\in K,~m\in M_k,~t\in T. \nonumber
  \end{alignat}
\end{minipage}
}

\medskip

Recall that, in this variant, we use the following notation: $\rho_{nk}=\frac{v_{nk}}{\sum_{m\in M_k}w_{nkm}d_{nkm}}$, $\xi_{nk}=\frac{\sum_{m\in M_k}w_{nkm}}{\min_{m\in M_k:w_{nkm}\ne0}w_{nkm}}$, $\rho_k^{\max}=\max_{n\in N}\rho_{nk}$, $\rho_k^{\min}=\min_{n\in N:\rho_{nk}>0}\rho_{nk}$, $\xi_k^{\max}=\max_{n\in N}\xi_{nk}$, $d_k^{\max}=\max_{n\in N,m\in M_k}d_{nkm}$, $d_k^{\min}=\min_{n\in N,m\in M_k:d_{nkm}>0}d_{nkm}$, $\bar \rho_k=\frac{\rho_k^{\max}}{\rho_k^{\min}}$, $\bar d_k=\frac{d_k^{\max}}{d_k^{\min}}$, $\bar\rho^{\max}=\max_{k\in K} \bar\rho_k$, $\xi^{\max}=\max_{k\in K} \xi_k^{\max}$ and $\bar d^{\max}=\max_{k\in K} \bar d_k$. In the algorithm, we choose  $\mu_{nkm} = e^{\frac{w_{nkm}\cdot\gamma_{k}}{C_{km}}}$ and $\beta_{nkm} =\rho^{\min}_k (\mu_{nkm}-1)$, for $n\in N$, $k\in K$, and $m\in M_{k}$, where
\begin{align}\label{gamma3}
\gamma_k=2\cdot\ln\left(4\cdot\bar\rho_k\cdot\bar d_k\cdot\xi_k^{\max}+2\right).
\end{align}

\begin{algorithm}[H]
  \caption{Primal-Dual Algorithm for \MdOMKD}
  \label{Alg_4}
  \begin{algorithmic}[1]
    \State Initialize \( \mathbf{x} \gets \mathbf{0} \), \( \mathbf{p} \gets \mathbf{0} \), \( \mathbf{u} \gets \mathbf{0} \)
    \For {each \( n \in N \)}
        \State Compute: \label{s3-1}
        \begin{align*}
            &k^* \gets \arg\max_{k \in K} \left\{ v_{nk} - \sum_{m\in M_k,t\in T_{nkm}}w_{nkm}p_{kmt}  \right\} 
        \end{align*}
        
        \If { \( \displaystyle  v_{nk^*} > \sum_{m\in M_{k^*},t\in T_{nk^*m}}w_{nk^*m}p_{k^*mt}\) }\label{s3-4}
            \State \( x_{nk^*} \gets 1 \)\label{s4-2}
            \State \( \displaystyle u_n \gets v_{nk^*} - \sum_{m\in M_{k^*},t\in T_{nk^*m}}w_{nk^*m}p_{k^*mt} \) \label{s3-6}
             \For{each \( m \in M_{k^*} \)}
            \For{each \( t \in T_{nk^*m} \)}
           
                \State
                \(p_{k^*mt} \gets \mu_{nk^*m} p_{k^*mt} + \beta_{nk^*m} \) 
            \EndFor\label{s3-9}
            \EndFor
        \EndIf
    \EndFor
  \end{algorithmic}
\end{algorithm}


\begin{theorem}
    Assuming that $\epsilon_{km}\leq \frac{C_{km}\ln 2}{\gamma_k}$, Algorithm \ref{Alg_4} outputs a feasible solution and has a competitive ratio of $O\big(\log(\bar \rho^{\max}\cdot\xi^{\max}\cdot\bar d^{\max})\big)$\label{thm_4}. 
\end{theorem}
To prove theorem \ref{thm_4}, we prove that the solutions obtained by Algorithm \ref{Alg_4} satisfy the following lemmas.


\begin{lemma}
(a) The dual solution generated by Algorithm \ref{Alg_4} is feasible.

(b) For all $k\in K$, $m\in M_k$ and $t\in T$, 
   $p_{kmt} =\rho^{\min}_k \left({e^{\frac{\gamma_k\cdot z_{kmt}}{C_{km}}}} - 1\right)$, 
where $z_{kmt}= \sum_{n\in N:t\in T_{nkm}} w_{nkm}x_{nk}$ is the {\it utilization} at time slot $t$ of dimension $m$ of resource $k$.\label{Lemma13}    
\end{lemma}
Let  $\Delta P^{\text{mmd}}$ and $\Delta D^{\text{mmd}}$ be the changes in the primal and dual objective values in one iteration, respectively. We say that a request $n\in N$ is admitted if $x_{nk^*}$ is set to $1$ in line~\ref{s4-2} of Algorithm \ref{Alg_4}. 
\begin{lemma}
(a) If an arriving request is not admitted, there is no change in the primal or the dual solution.\\
(b) If an arriving request is admitted, $\frac{\Delta D^{\text{mmd}}}{\Delta P^{\text{mmd}}}$ is bounded by 
 $\frac{2\gamma_{k^*}}{\ln 2}$. \label{Lemma14}
\end{lemma}
\begin{lemma}
The primal solution generated by  Algorithm \ref{Alg_4} is feasible.\label{Lemma15}
\end{lemma}

\sproofof{Lemma~\ref{Lemma13}}  
The proof is similar to that of Lemma~\ref{Lemma1} and is omitted. \qed
\hide{
(b) Initially, $z_{kmt}=0$ and ${p}_{kmt}=0$. By the update formula, \(p_{mk^*t} \gets \mu_{nmk^*} p_{mk^*t} + \beta_{nmk^*} \). Also, let $\bar{z}_{k^*tm} =z_{k^*tm} + w_{nk^*m}$ be the utilization after admission of request $n$. After admission, the updating formula in Algorithm \ref{Alg_4} becomes
\begin{align}
  \bar{p}_{mkt} &= \mu_{nk^*m}\rho^{\min}_k\left(e^{\frac{\gamma_{k} z_{ktm}}{C_{km}}}-1 \right)  + \beta_{nkm}\nonumber\\
  &=\mu_{nkm}\rho^{\min}_k\left(e^{\frac{\gamma_{k} z_{ktm}}{C_{km}}}-1 \right) + \rho^{\min}_k(\mu_{nkm}-1)\nonumber\\
  &= e^{\frac{w_{nkm}\gamma_{k}}{C_{km}}} \left(e^{\frac{\gamma_{k} z_{k^*tm}}{C_{km}}}-1 \right) + \left(e^{\frac{w_{nkm}\gamma_{k}}{C_{km}}}-1 \right)\nonumber\\
  &= \rho^{\min}_k\left(e^{\frac{\gamma_{k} \bar{z}_{ktm}}{C_{km}}}-1\right). \label{A_Accept}
\end{align}
}

\medskip

\sproofof{Lemma~\ref{Lemma14}} (a) If a request is rejected on arrival by Algorithm \ref{Alg_4}, then obviously $\Delta P^{\text{mmd}}=\Delta D^{\text{mmd}}=0$.

    (b) Suppose that a new request $n$ is admitted to resource $k^*$ in an iteration, then the change in the primal objective value is
    \begin{equation}
        \Delta P^{\text{mmd}}= v_{nk^*}\label{Delta_p_mm},
    \end{equation}
     and the change in the dual objective value is
    \begin{equation}
        \Delta D^{\text{mmd}} = \sum_{m\in M_{k^*}, t\in T_{nk^*m}}C_{k^*m}\Delta p_{k^*mt} + u_n.\label{delta_mmd}
    \end{equation}

     By substituting $u_n=v_{nk^*} - \sum_{m\in M_{k^*},t\in T_{nk^*m}}w_{nk^*m}p_{k^*mt}$ from Algorithm \ref{Alg_4} into (\ref{delta_mmd}), we obtain

{\small
\begin{align}\label{Delta_d_mm}
    \Delta D^{\text{mmd}} &= \sum_{m\in M_{k^*}, t\in T_{nk^*m}}C_{k^*m}(\mu_{nk^*m} p_{k^*mt} + \beta_{nk^*m}-p_{k^*mt}) +  v_{nk^*} - \sum_{m\in M_{k^*}, t\in T_{nk^*m}}w_{nk^*m}p_{k^*mt}\nonumber\\
    & = v_{nk^*} +\sum_{m\in M_{k^*}, t\in T_{nk^*m}}C_{k^*m}[(\mu_{nk^*m}-1)-w_{nk^*m}] p_{k^*mt} +   \sum_{m\in M_{k^*}}C_{k^*m}\beta_{nk^*m}d_{nk^*m}\nonumber\\
    &\le v_{nk^*} +\left(\frac{\gamma_{k^*}}{\ln 2}-1\right)\sum_{m\in M_{k^*}, t\in T_{nk^*m}}w_{nk^*m}p_{k^*mt}+\frac{\gamma_{k^*}\cdot\rho_{k^*}^{\min}}{\ln 2} \sum_{m\in M_{k^*}}w_{nk^*m}d_{nk^*m}\nonumber\\
    &\le \frac{\gamma_{k^*}}{\ln 2}\left(v_{nk^*}+\rho_{k^*}^{\min}\sum_{m\in M_{k^*}} w_{nk^*m}  d_{nk^*m}\right),
\end{align}
}

\noindent where the first inequality follows from the fact that $C_{k^*m}[(\mu_{nk^*m}-1)-w_{nk^*m}]\le \left(\frac{\gamma_{k^*}}{\ln 2}-1\right)w_{nk^*m}$ (which holds by our assumption that $w_{nk^*m}\le\frac{C_{k^*m}\ln 2}{\gamma_{k^*}}$) and the second inequality follows from the admission policy in line \ref{s1-4} of Algorithm \ref{Alg_4}.

\medskip

It follows from~\raf{Delta_p_mm} and~\raf{Delta_d_mm} that
\begin{align*}
  \frac{\Delta D^{\text{mmd}}}{\Delta P^{\text{mmd}}}   &\leq \frac{\gamma_{k^*}}{\ln 2}\left(1+\rho_{k^*}^{\min}\sum_{m\in M_{k^*}} \frac{w_{nk^*m}  d_{nk^*m}}{v_{nk^*}}\right)=\frac{\gamma_{k^*}}{\ln 2}\left(1+\frac{\rho_{k^*}^{\min}}{\rho_{nk^*}}\right)\le \frac{2\gamma_{k^*}}{\ln 2}.
\end{align*}
The lemma follows.
\qed

\medskip

\sproofof{Lemma~\ref{Lemma15}} In order to show that the primal solution generated by Algorithm \ref{Alg_4} is feasible, we need to show that once the capacity of a resource is full, the resource stops accepting further requests. For the sake of a contradiction, suppose that there is a $t_0\in T$ and $k^*\in K$ such that a violation happens at time $t_0$ for resource $k^*$ in dimension $m^*\in M_{k^*}$, i.e., we have  $\sum_{n\in N:t_0 \in T_{nk^*m^*}}w_{nk^*m^*}x_{k^*n}>C_{k^*m^*}$.
    As in the proof of Lemma~\ref{Lemma3}, we can argue that there exists a request $n_0$ with strictly positive weight $w_{n_0k^*m^*}$ that got admitted to resource $k^*$, and a subset of time slots $T_0\subseteq T_{n_0k^*m^*}$ of length at least $\frac{d^{\min}_{k^*}}2$, such that, prior to admitting $n_0$, $z_{k^*m^*t}\ge\frac{C_{k^*m^*}}{2} - \frac{C_{k^*m^*} \ln 2}{\gamma}$.
Then we get from Lemma~\ref{Lemma13}(b) that 
\begin{align}\label{e5}
   \sum_{t\in T_{n_0k^*m^*}}p_{k^*m^*t} &= \rho^{\min}_{k^*}\sum_{t \in T_{n_0k^*m^*}} \left( e^{\frac{\gamma_{k^*}\cdot z_{k^*m^*t}}{C_{k^*m^*}}} - 1  \right) \geq \rho^{\min}_{k^*}\sum_{t \in T_0} \left( e^{\frac{\gamma_{k^*}\cdot z_{k^*m^*t}}{C_{k^*m^*}}} - 1  \right)\nonumber\\
   &\geq \rho^{\min}_{k^*}\sum_{t \in T_0}  \left( e^{\frac{\gamma_{k^*} \left(\frac{C_{k^*m^*}}{2} - \frac{C_{k^*m^*} \ln 2}{\gamma_{k^*}}\right)}{C_{k^*m^*}}} - 1  \right)\nonumber\\
   &=|T_0|\rho^{\min}_{k^*} \left( e^{\left(\frac{\gamma_{k^*}}{2} -  \ln 2\right)} - 1  \right)\nonumber\\ &\ge \frac{d^{\min}_{k^*}\cdot\rho^{\min}_{k^*}}2 \left( e^{\left(\frac{\gamma_{k^*} }{2} -  \ln 2\right)} - 1  \right).
\end{align}
By definition~\raf{gamma3} of $\gamma_{k^*}$,  
\begin{align}
\gamma_{k^*} &=2\ln\left(4\cdot\frac{\rho^{max}_{k^*}}{\rho^{\min}_{k^*}}\cdot\frac{d^{\max}_{k^*}}{d^{\min}_{k^*}}\cdot\xi^{\max}_{k^*}+2\right)\nonumber\\
&\geq 2 \ln \left(4\cdot\frac{v_{n_0k^*}}{\sum_{m\in M_{k^*}}w_{n_0k^*m}d_{n_0k^*m}\cdot\rho^{\min}_{k^*}}\cdot\frac{\sum_{m\in M_{k^*}}w_{n_0k^*m}d_{n_0k^*m}}{\sum_{m\in M_{k^*}}w_{n_0k^*m}\cdot d^{\min}_{k^*}}\cdot\frac{\sum_{m\in M_{k^*}}w_{n_0k^*m}}{w_{n_0k^*m^*}} +2\right)\nonumber\\ 
&\ge2\ln\left(2+4\cdot\frac{v_{n_0k^*}}{w_{n_0k^*m^*}}\cdot\frac1{\rho^{\min}_{k^*}\cdot d^{\min}_{k^*}}\right)
 \label{lower_bound_gamma_MKD}.
\end{align}
Combining (\ref{e5}) and (\ref{lower_bound_gamma_MKD}) we obtain
\begin{align}
 \sum_{t \in T_{n_0k^*m^*}} p_{k^*m^*t} &> \frac{d^{min}_{k^*}\cdot\rho^{\min}_{k^*}}{2} \left( \frac{2 v_{n_0k^*}}{w_{n_0k^*m^*}}\cdot\frac1{\rho^{\min}_{k^*}\cdot d^{\min}_{k^*}} \right) = \frac{v_{n_0k^*}}{w_{n_0k^*m^*}},
\end{align}
which implies that $\sum_{m\in M_{k^*},t \in T_{n_0k^*m}} w_{n_0k^*m}p_{k^*mt}\ge w_{n_0k^*m^*}\sum_{t\in T_{n_0k^*m^*}}p_{k^*m^*t}>v_{n_0k^*}$, leading to the contradiction that $n_0$ was not admitted, according to our admission policy in line~\ref{s3-4} of the algorithm.
\qed

\bibliographystyle{elsarticle-num}


\end{document}